\documentclass[twocolumn, showpacs, showkeys, prb, floatfix, 
superscriptaddress]{revtex4}
\usepackage{graphicx}
\usepackage{bm}
\usepackage{amsmath}
\usepackage{amssymb}

\newcommand* {\etal}{\emph{et al.}}
\DeclareMathOperator{\hrmcnjg}{hc}
\DeclareMathOperator{\trace}{tr}
\newcommand* {\vek}[1]{{\bm{\mathrm{#1}}}}
\newcommand* {\vekc}[1]{{\bm{\mathcal{#1}}}}
\newcommand* {\kk}{\vek{k}}
\newcommand* {\rr}{\vek{r}}
\newcommand* {\pp}{\vek{p}}
\newcommand* {\kdotp}{\kk\cdot\pp}
\newcommand* {\frack}[2]{{\textstyle\frac{#1}{#2}}}
\newcommand* {\bra}[1]{\langle {#1} |}
\newcommand* {\ket}[1]{| {#1} \rangle}
\newcommand* {\braket}[1]{\langle {#1} \rangle}
\usepackage{array}
\newcolumntype {s}[1]{@{\hspace{#1}}} 

\begin{document}
\title{Electron spin orientation under in-plane optical excitation
 in GaAs quantum wells}

\author{S. Pfalz}
\affiliation{Institut f\"{u}r
Festk\"{o}rperphysik, Gottfried Wilhelm Leibniz Universit\"{a}t
Hannover, Appelstra{\ss}e 2, D-30167 Hannover, Germany}

\author{R. Winkler}
\affiliation{Department of Physics, Northern Illinois University,
DeKalb, IL 60115, USA}
\affiliation{Department of Physical Chemistry, 
 The University of the Basque Country, 48080 Bilbao, Spain}
\affiliation{IKERBASQUE, Basque Foundation for Science, 48011
  Bilbao, Spain}

\author{N. Ubbelohde}
\affiliation{Institut f\"{u}r Festk\"{o}rperphysik, Gottfried
Wilhelm Leibniz Universit\"{a}t Hannover, Appelstra{\ss}e 2,
D-30167 Hannover, Germany}

\author{D. H\"{a}gele}
\affiliation{Spectroscopy of Condensed Matter,
Ruhr-Universit{\"a}t Bochum, 44801 Bochum, Germany}

\author{M. Oestreich}
\affiliation{Institut f\"{u}r Festk\"{o}rperphysik, Gottfried
Wilhelm Leibniz Universit\"{a}t Hannover, Appelstra{\ss}e 2,
D-30167 Hannover, Germany}

\date{8 August 2012}

\begin{abstract}
  We study the optical orientation of electron spins in GaAs/AlGaAs
  quantum wells for excitation in the growth direction and for
  in-plane excitation. Time- and polarization-resolved
  photoluminescence excitation measurements show, for resonant
  excitation of the heavy-hole conduction band transition, a
  negligible degree of electron spin polarization for in-plane
  excitation and nearly 100\% for excitation in the growth
  direction. For resonant excitation of the light-hole conduction
  band transition, the excited electron spin polarization has the
  same (opposite) direction for in-plane excitation (in the growth
  direction) as for excitation into the continuum. The experimental
  results are well explained by an accurate multiband theory of
  excitonic absorption taking fully into account electron-hole
  Coulomb correlations and heavy-hole light-hole coupling.
\end{abstract}

\maketitle


\section{Introduction}
\label{sec:intro}

The efficient injection and detection of spin-polarized carriers in
semiconductor quantum wells (QWs) is a field of intensive research.
\cite{zut04} Among various approaches that have been used to achieve
this goal, optical selection rules \cite{dya84} have often played an
important role. In many cases, spin-polarized electrons are both
optically excited and optically detected. \cite{hae98, cro05, gil09}
Other approaches combine optical excitation of spin-polarized
carriers with electrical detection schemes utilizing, e.g.,
orientation- and spin-dependent charge currents. \cite{gan03a} Still
others employ electrical injection of spin-polarized carriers, e.g.,
via paramagnetic semiconductors as spin aligners, and probe
optically by polarization-resolved photoluminescence (PL)
spectroscopy. \cite{oes99, fie99, jon00} All these experiments
demonstrate that optical selection rules are a very useful tool
to study semiconductor spintronics since they directly relate the
light polarization with the spin polarization of the electrons.
\cite{dya84} In principle, the optical selection rules also
determine the hole spin polarization in the valence band, yet this
polarization usually decays so rapidly \cite{dam91, hil02} that it can
be neglected in most experiments. Nonetheless, recent experiments
have also exploited the optical orientation for hole systems.
\cite{kor10, pri10a}

The optical selection rules in QWs have been investigated in detail
for optical excitation by circularly polarized light in the
\emph{growth direction}, \cite{pfa05} but the selection rules for
spin excitation \emph{in the plane} of the QW have not been studied
systematically so far, to the best of our knowledge. Such an in-plane
excitation or detection of electron spin polarization plays an
important role in a variety of experiments. For example, Ohno \etal\
measured the degree of circular polarization of the side-emitted
electroluminescence due to the heavy-hole (HH) transition of electrically
injected carriers. \cite{ohn99a, oes99a} Oest\-reich \etal\ studied
spin precession in a magnetic field after in-plane excitation of the
light-hole (LH) transition to directly measure the sign of the effective
$g$ factor of electrons in a QW. \cite{oes98}

In this article, we present a detailed study of the optical
orientation of electron spins in a GaAs multi-QW using a light beam
propagating \emph{parallel} to the plane of the two-dimensional (2D)
system. A circularly polarized laser pulse is focused on the cleaved
edge of the QWs, creating spin-polarized electrons in the wells.
Application of an in-plane magnetic field perpendicular to the
excitation direction leads to spin precession, which we observe in
the optical emission in the growth direction of the 2D system. From
the time- and polarization-resolved PL, we obtain the
initial degree of electron spin polarization $P_0$ which is studied
as a function of the excitation energy. We compare our measured
results with an accurate theory of excitonic absorption taking fully
into account electron-hole Coulomb correlations and HH-LH coupling.

We begin in Sec.~\ref{sec:qual} by discussing a simplified,
qualitative model that incorporates the main features of optical
orientation for arbitrary excitation and polarization directions.
The experimental setup and methods for data analysis are described
in Sec.~\ref{sec:methods}. Section~\ref{sec:Res_Growth} presents as
a reference frame the results for excitation in the growth direction,
while the rest of Sec.~\ref{sec:results} is devoted to in-plane
excitation. The accurate theoretical model is presented in
Sec.~\ref{sec:theory}. We end with conclusions in
Sec.~\ref{sec:Conclusions}.

\section{Qualitative Model for Optical Orientation}
\label{sec:qual}

For QWs made of direct semiconductors such as GaAs, the main
features of optical orientation can be understood in a simplified
version of the full theory developed in Sec.~\ref{sec:theory}. In
this simplified model, we neglect the in-plane dispersion of the
electron and hole states (i.e., in-plane wave vector $\kk = 0$) as
well as the $\kdotp$ coupling between conduction and valence band
states so that the subband states are represented by their dominant
spinor components. If the exciting light is characterized by a
polarization vector $\hat{\vek{e}} = (e_x,e_y,e_z)$, the optically
excited electrons are characterized by the $2 \times 2$ spin density
matrix \cite{scalar}
\begin{equation}
  \label{eq:spinmat-simple}
  \dot{\varrho} (\omega)
  \equiv \frac{d}{dt} \, \rho (\omega)
  = \mathcal{C} \sum_\alpha
  \left( \varrho_0^\alpha \openone_{2\times 2} +
    \vek{s}^\alpha \cdot \vek{\sigma} \right)
  \delta(\hbar\omega - E_\alpha) ,
\end{equation}
where the sum runs over HH and LH states, and $\vek{\sigma}$ is the
vector of Pauli matrices. We have, for the HH transitions,
\begin{subequations}
  \begin{eqnarray}
    \varrho_0^\mathrm{HH} & = & |e_x|^2 + |e_y|^2 \\
    s_x^\mathrm{HH} & = & s_y^\mathrm{HH} = 0 \\
    s_z^\mathrm{HH} & = & 2 \, \Im \left( e_x e_y^\ast\right)
  \end{eqnarray}
\end{subequations}
and for the LH transitions
\begin{subequations}
  \begin{eqnarray}
    \varrho_0^\mathrm{LH} & = &
    \frack{1}{3}|e_x|^2 + \frack{1}{3}|e_y|^2
    + \frack{4}{3} |e_z|^2 \label{eq:dens_lh_0}\\
    s_x & = & \frack{4}{3} \, \Im \left( e_y e_z^\ast\right) \\
    s_y & = & \frack{4}{3} \, \Im \left( e_z e_x^\ast\right) \\
    s_z & = & - \frack{2}{3} \, \Im \left( e_x e_y^\ast\right) ,
    \hspace{2em}
  \end{eqnarray}
\end{subequations}
which reflects the well-known Clebsch-Gordan coefficients
characterizing the dipole matrix elements between spin-$1/2$ states
in the conduction band and (effective) spin-$3/2$ states in the
valence band. \cite{dya84} Finally, we have
\begin{equation}
  \mathcal{C} = \frac{e^2 A^2_0}{2m_0} \frac{P^2}{\hbar\omega} ,
\end{equation}
where $A_0$ denotes the amplitude of the vector potential of the
light field and $P$ is Kane's momentum matrix element. \cite{win03}
The quantity $\mathcal{C}$ is essentially a constant for optical
transitions close to the fundamental absorption edge. We see that,
apart from the prefactor $\mathcal{C}$ the spin density matrix
depends only on the components of the polarization vector
$\hat{\vek{e}}$ and the excitation energies $E_\mathrm{HH}$ and
$E_\mathrm{LH}$ for HH and LH transitions. (We have ignored the
trivial selection rule that we get a large oscillator strength for
optical transitions only if the envelope functions for the electron
and hole state have the same number of nodes.) Similar results for
bulk material were previously obtained by Dymnikov \etal
\cite{dym76}

Apart from a constant prefactor [see Eq.\ (\ref{eq:absorb}) below],
the absorption coefficient is then given by
\begin{subequations}
  \label{eq:spinpol-simple}
  \begin{equation}
    \alpha (\omega) \propto \trace \dot{\varrho}
    = 2 \mathcal{C} \sum_\alpha
    \varrho_0^\alpha \, \delta(\hbar\omega - E_\alpha) ,
  \end{equation}
  where $\trace$ denotes the trace, and the spin polarization
  induced by a steady-state optical excitation becomes
  \begin{equation}
    \vek{P} (\omega)
    = \frac{\trace \vek{\sigma} \dot{\varrho}}{\trace \dot{\varrho}}
    = \left\{ \begin{array}{cs{2em}l}
        \vek{s}^\mathrm{HH} / \varrho_0^\mathrm{HH}, &
        \hbar\omega=E_\mathrm{HH} \\[1ex]
        \vek{s}^\mathrm{LH} / \varrho_0^\mathrm{LH}, &
        \hbar\omega=E_\mathrm{LH} .
      \end{array} \right.
  \end{equation}
\end{subequations}
Equations (\ref{eq:spinpol-simple}) can be easily evaluated for
different polarization vectors $\hat{\vek{e}}$. They include the
well-known result \cite{dya84, pfa05} that, for excitation in the growth
direction, the HH transitions are three times more efficient than the
LH transitions (independendent of the polarization $\hat{\vek{e}}$).
With circularly polarized light [$\hat{\vek{e}} =
\frack{1}{\sqrt{2}} (1,\pm i,0)$] we obtain for both HH and LH
transitions a complete electron spin polarization ($|\vek{P}|= 1$)
in the $z$ direction that is opposite in sign for HH and LH transitions.
It also follows from these equations that an in-plane excitation can
give rise to both HH and LH transitions. Yet the details depend on
the polarization $\hat{\vek{e}}$. HH states do not couple to $e_z$
from which it follows immediately that an electron spin polarization
via in-plane excitation of HH transitions is not possible.
\cite{fie03, oes05} Only LH transitions with circularly polarized
light [e.g., $\hat{\vek{e}} = \frack{1}{\sqrt{2}} (1,0,\pm i)$] can
give rise to a spin polarization of the electron states with a
maximum $|\vek{P}| = 4/5$.

While Eqs.\ (\ref{eq:spinmat-simple}) and (\ref{eq:spinpol-simple})
allow one to understand the main features of optical orientation in
QWs, these results get modified by the details of band structure
such as HH-LH coupling [Eq.\ (\ref{eq:spinmat-single}) below] and
the excitonic Coulomb coupling between the single-particle states
[Eq.\ (\ref{eq:spinmat-ex})].

\section{Experimental Methods}
\label{sec:methods}

Our sample is a multi-QW structure consisting of $15$ layers of
GaAs/Al$_{0.3}$Ga$_{0.7}$As QWs with a well width of $14$~nm
separated by barriers with a thickness of $10$~nm. The structure is
grown on a $(001)$-oriented GaAs substrate and sandwiched between
layers of 500~nm and 490~nm Al$_{0.3}$Ga$_{0.7}$As. The sample is of
a very high quality, which has been confirmed by absorption
measurements [see, e.g., the left column in
Fig.~\ref{fig:Growth_InPlane}(c) and measurements on a very similar
sample containing 10 instead of 15 QWs in
Ref.~\onlinecite{loe02}]. The Stokes shift is, at most, of the order
of $0.3$~meV, which is the resolution limit of the experimental
setup.

\begin{figure}[t]
\includegraphics{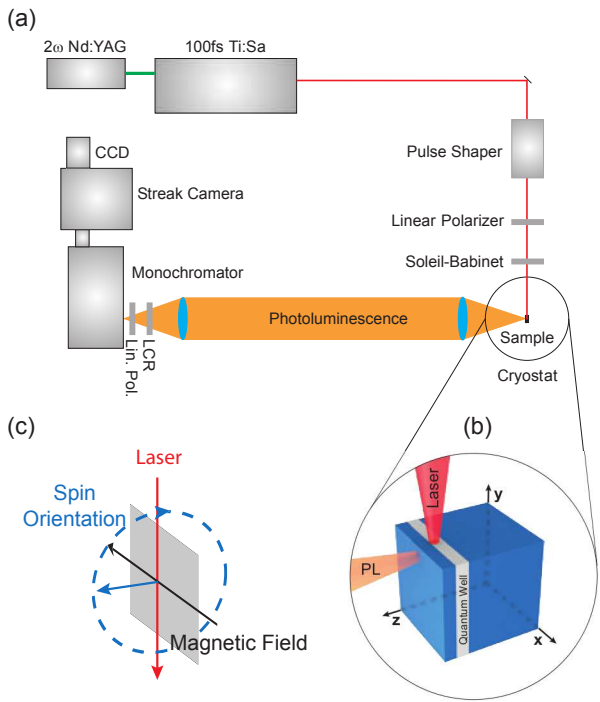}
\caption{\label{fig:setup}(Color online) (a) Experimental setup of
the time-resolved PL measurements, (b) geometry for laser
excitation as well as PL detection, and (c) sketch of the
precession of the spin orientation (blue arrow) around the
applied in-plane magnetic field (black arrow).}
\end{figure}

In the following, we present measurements of the initial degree of
spin polarization for two geometries of optical
excitation. First, we perform a control experiment, where the sample
is excited with circularly polarized laser pulses in the growth
direction. We label the growth direction of the sample the $z$ axis,
while the $x$ and $y$ directions are oriented in the plane of the QW
[see Fig.~\ref{fig:setup}(c)]. The experimental setup for this
measurement as well as the techniques for data analysis is
described in detail in Ref.~\onlinecite{pfa05}. Second, we focus the
circularly polarized laser pulses on the cleaved edge of the sample
and excite electrons with a spin polarization in the plane of the QW
($y$ direction). The setup for the latter experiments is described
in detail in the following.

Figure~\ref{fig:setup}(a) depicts the experimental setup for the
time- and polarization-resolved PL measurements. The sample is
mounted in Voigt configuration in a He gas flow cryostat in a
superconducting magnet, cooled to a lattice temperature of 10~K, and
excited by pulses from a femtosecond mode-locked Ti:sapphire laser with a
repetition rate of $80$~MHz. A pulse shaper reduces the spectral
linewidth of the $100$~fs laser pulses to $0.8$~nm full width at
half-maximum (FWHM) and a Soleil-Babinet compensator adjusts the
polarization of the laser pulses at the sample surface to circular
polarization. Unless stated otherwise, the time-averaged excitation
power is $1$~mW. The exciting laser light is propagating in the $y$
direction. An in-plane magnetic field $B_x$ gives rise to Larmor
precession of the electron spins around $B_x$ [also called spin
quantum beats (SQB), see Fig~\ref{fig:setup}(c)] with the Larmor
frequency $\omega_L = g^\ast \mu_B B_x / \hbar$, where $g^\ast$ is
the effective electron $g$ factor, $\hbar$ is the Planck constant,
and $\mu_B$ is the Bohr magneton. We detect the PL components along
the $z$ axis, i.e., we detect the projection of the electron spin
orientation on the $z$ axis which oscillates with the frequency
$\omega_L$. For this purpose, the intensities of the left and right
circularly polarized PL components $I_{\pm}(t)$ are measured
separately using an electrically tunable liquid-crystal retarder
(LCR), a linear polarizer, and a synchroscan streak camera providing
temporal and spectral resolutions of $15$~ps and $7$~meV,
respectively. The resulting time-dependent degree of optical
polarization is defined as
\begin{equation}
  \label{eq:Popt}
  P_\mathrm{opt}^z (t) = \frac{I_+(t) - I_-(t)}{I_+(t) + I_-(t)}.
\end{equation}
At $t=0$, the electron spins are initially oriented along the $y$
axis so that the measured $P_\mathrm{opt}^z (t=0)$ is zero. After
$1/4$ of the oscillation period, $P_\mathrm{opt}^z$ has either a
maximum or a minimum, depending on the magnetic field direction and
the sign of $g^\ast$. \cite{oes98} In order to determine the initial
electron spin polarization $P_0$ at $t=0$, the measured
$P_\mathrm{opt}^z (t)$ is fitted by the expression
\begin{equation}
  \label{eq:extrapol}
  P_\mathrm{opt}^z (t) = P_0 \,
  \exp\left(- t/\tau_s\right)
  \sin\left(\omega_L t\right)
  + p_0 ,
\end{equation}
where $\tau_s$ is the electron spin relaxation time, and $p_0$ is a
systematic offset in our measurements which results from the liquid
crystal retarder. In most of our fits we find $\left|p_0\right| \leq
0.02$.

Equation (\ref{eq:extrapol}) is based on three assumptions, which are
discussed in detail in Sec.~II of Ref.~\onlinecite{pfa05} and
references therein. First, hole spin relaxation is assumed to be
fast compared to electron spin relaxation. This assumption is
supported by several experiments and calculations which show that
the spin relaxation of free holes is of the order of the momentum
relaxation time. \cite{hil02, dam91} Second, the measured PL results
solely from recombination of the HH1:E1 transition, and HH-LH mixing
can be neglected in that case. This assumption is supported by our
measurements since we find that $P_\mathrm{opt}^z (0)$ is close to
100\% for resonant, circularly polarized excitation of the HH
transition in the growth direction. Third, the electron spin relaxation
is mono-exponential which is validated by all our fits.

\begin{figure}[t]
\includegraphics{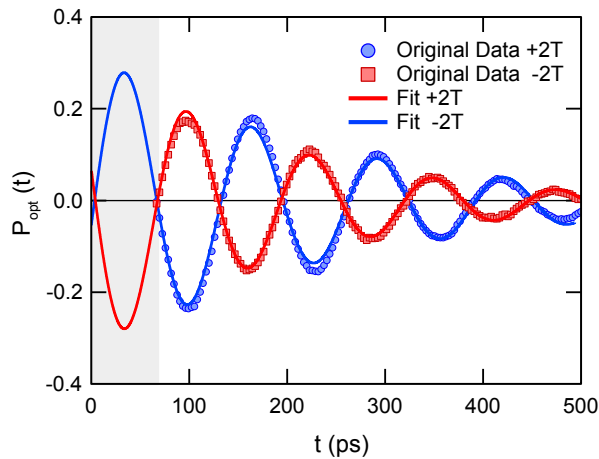}
\caption{\label{fig:SQB}(Color online) Typical experimental data and
 fits for an excitation energy of $1.567$~eV and an excitation
 intensity of 6~mW. The time of laser excitation defines $t=0$. The
 red squares (blue circles) show the measured SQBs with an applied
 magnetic field of $+2$~T ($-2$~T) while the dashed lines represent
 the corresponding fits. Within the blue shaded area, laser
 straylight may influence the measured degree of polarization.
 Therefore, the data are fitted only outside the blue shaded area
 for ($t-t_0 \ge 80$~ps).}
\end{figure}

Figure~\ref{fig:SQB} shows an SQB measurement for two
magnetic fields of $+2$~T (red squares) and $-2$~T (blue circles).
The measured SQBs are shown for $t \ge 80$~ps since, in this geometry,
laser stray light obstructs the detection of SQBs during the first
picoseconds after the excitation. Depending on the excitation energy
this time frame varies and it may last up to 80~ps for excitation at
the HH resonance. The dashed lines depict the fits of the SQBs
according to Eq.\ (\ref{eq:extrapol}). The fits clearly yield
$P_\mathrm{opt}^z (t=0) = 0$, i.e., the spin excitation is solely
in-plane. This is an important consistency check to rule out
unintentional excitation in the growth direction. Such an unintentional
excitation could occur if part of the exciting laser light hits the
growth surface and this light is then diffracted into the growth
direction due to the large refractive index of GaAs.

It is conceivable that fast spin relaxation mechanisms effective at
early time scales not accessible in our measurements may affect the
spin polarization measured by extrapolation to time $t=0$. However,
in our experiments it appears unlikely that such a mechanism plays a
significant role because any such additional spin relaxation channel
must further reduce the measured degree of spin polarization. The
overall good agreement of the absolute values of the measured versus the
calculated spin orientations as a function of the laser energy [see
Sec.~\ref{sec:Res_InPlane} and Figs.~\ref{fig:Growth_InPlane}(a) and
(b)] suggests that, overall, such an additional spin relaxation
channel is not important. The particular case of the LH1:E1
resonance is discussed in more detail in Sec.~\ref{sec:Bleaching}.

\section{Results and Discussion}
\label{sec:results}

\subsection{Excitation in the growth direction}
\label{sec:Res_Growth}

\begin{figure*}[t]
\includegraphics[width=0.85 \linewidth]{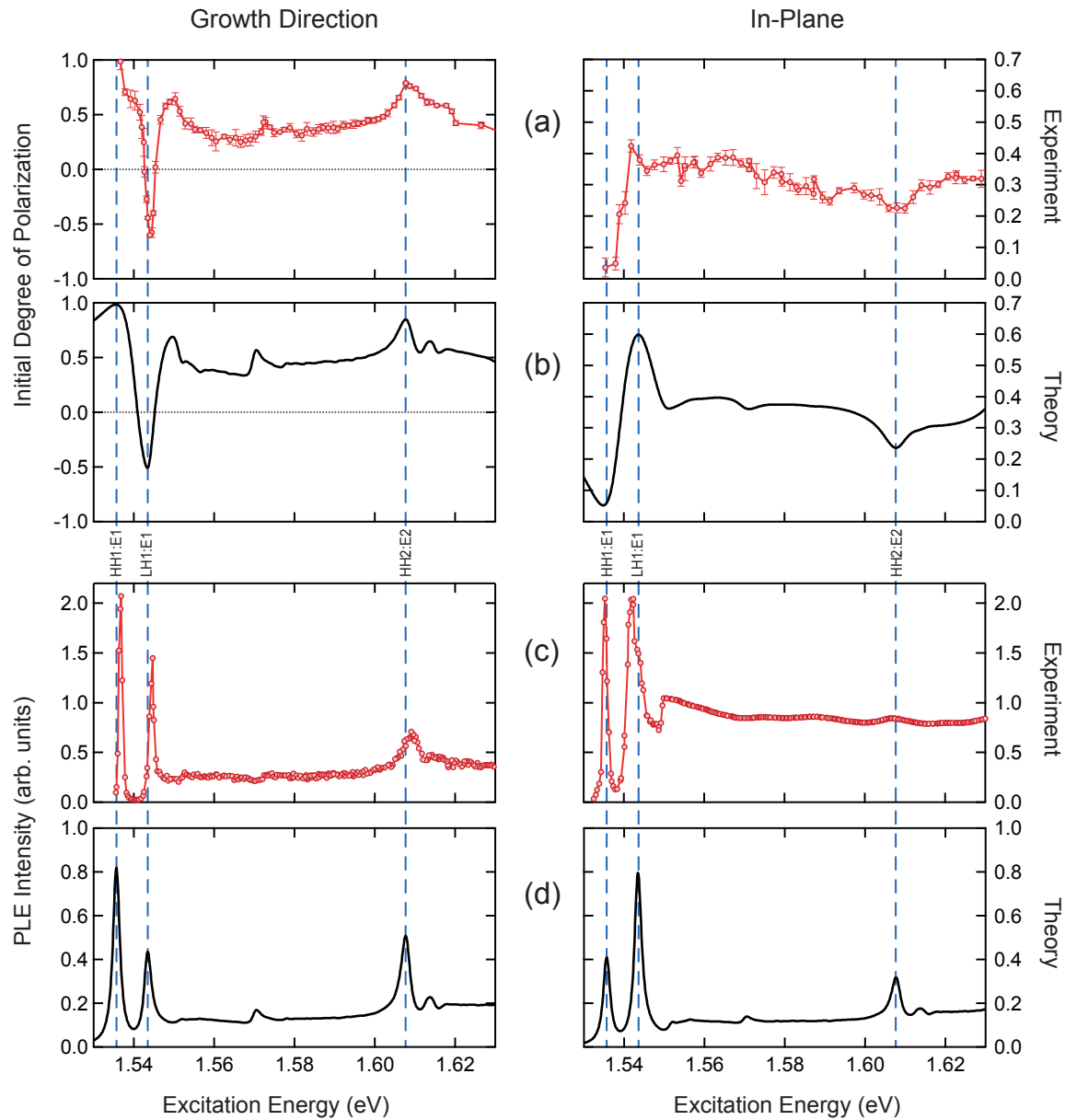}
\caption{\label{fig:Growth_InPlane}(Color online) Comparison of (a)
 the experimental results and (b) calculated values for the initial
 degree of electron spin polarization as a function of the
 excitation energy. The left (right) column shows the data for
 excitation in growth-direction (in the QW-plane). The lower panels
 show the absorption coefficient obtained from PLE- measurements (c)
 and theory (d) in growth direction (left) and for in-plane
 excitation (right).}
\end{figure*}

We perform measurements of $P_0$ for excitation in the growth direction
to characterize the sample and for comparison with the results
obtained for in-plane excitation. The left column in
Fig.~\ref{fig:Growth_InPlane} shows (a) the measured $P_0$, (b) the
calculated $P_0$, and (c) the measured PL excitation (PLE) spectrum
as a function of the excitation energy. We extract from the PLE
spectrum a full width at half maximum (FWHM) of the lowest HH
transition of $1$~meV and use this value for a phenomenological
Lorentzian broadening of the numerically calculated spectrum. We
briefly discuss the features of the measured $P_0$ going from low to
high excitation energies. For resonant excitation at $1.537$~eV of
the transition from the first HH subband to the first electron
subband (HH1:E1), we find, as expected, $P_0 = 1$. Around the LH1:E1
transition at $1.544$~eV, we observe a sign reversal of $P_0$
leading to a maximum negative initial degree of spin polarization of
$P_0 \approx -0.6$. The next peak at $1.55$~eV reflects the
absorption edge of the HH1:E1 exciton continuum. We find two
additional peaks at $1.572$~eV and $1.61$~eV, which correspond to
the HH3:E1 and HH2:E2 transition, respectively. All experimental
features are well reproduced by the calculated spectra.

\subsection{In-plane excitation}
\label{sec:Res_InPlane}

The right column in Fig.~\ref{fig:Growth_InPlane} shows (a) the
measured $P_0$, (b) the calculated $P_0$, (c) the measured PLE, and
(d) the calculated PLE as a function of the excitation energy. We use
for these calculations a Lorentzian broadening with an FWHM of 3~meV
which is consistent with the measured PLE spectrum for in-plane
excitation.

The significantly larger broadening for in-plane excitation results
probably from surface effects such as local oxidation of the AlGaAs
QW barriers at the cleaved surface. Moreover, our PLE measurements
are noisier for in-plane excitation than for excitation in the growth
direction since the PL intensity is lower. Finally, we note that we
have to use a smaller laser-spot diameter for in-plane excitation so
that small changes of the position of the exciting laser spot
strongly change the detected PL intensity. Nonetheless, we find no
indication that the measured PL signal contained contributions from
the LH1:E1 transition, consistent with the fact that this transition
lies about 7~meV above the HH1:E1 transition, which makes such a
contribution rather unlikely.

All features in the spectrum for in-plane excitation are
approximately 2~meV lower in energy than the corresponding features
obtained for excitation in the growth direction. We expect that this
result was caused by the position-dependent inhomogeneous strain that
was unintenionally present in the sample during the low-temperature
experiments, thus resulting in a small shift of the resonance
energies. \cite{jag86} The sample was glued to a sample holder such
that the edge used for the in-plane excitation was free-standing.
Thus upon cooldown the edge might have experienced a different
strain compared to the rest of the sample. We note, furthermore,
that the spectra for in-plane excitation and for excitation in
growth direction were measured in different cooldowns, which likewise
could have resulted in different amounts of unintentional strain in
these experiments.

In the PLE spectrum measured for in-plane excitation the peak
attributed to the LH1:E1 exciton has an asymmetric line shape,
suggesting that two excitons with almost the same energy contribute
to this peak. Such a doublet structure may occur if the standard
selection rules for a symmetric QW are relaxed due to the
presence of a symmetry-breaking perpendicular electric field.
\cite{win95a} We found that for the system studied here a weak field
of a few kV/cm gives rise to a second resonance slightly above the
LH1:E1 resonance. For our rather sensitive setup we cannot exclude
that such an electric field is present at the cleaved surface. We
note that our calculations indicate that the additional resonance
does not significantly affect the measured initial spin polarization
$P_0$. The features in the spectra at higher energies are not
affected either by such a weak electric field.

Again, we discuss the features of $P_0$ going from low to high
excitation energy. For resonant excitation of the HH1:E1
transition at $1.535$~eV, we measure $P_0 \approx 0.04$, which is close to
the calculated $P_0$ [see Fig.~\ref{fig:Growth_InPlane}(b)]. Our
calculations show that this small but finite $P_0$ results from the
broadening of the LH1:E1 transition and that $P_0$ vanishes with
decreasing broadening. This is consistent with the simplified model
in Sec.~\ref{sec:qual} which suggests that the polarization of the
HH1:E1 PL in $x$ direction is always linearly polarized independent
of the electron spin polarization, i.e., the PL emitted in-plane of
a QW gives no indication about the electron spin polarization in the
case of the HH1:E1 transition. This would be different if the LH
state contributed to the PL transition, e.g., if LH and HH
transitions overlapped due to broadening, if the LH states were
thermally occupied, or if the LH transition were energetically below
the HH transition due to strain. \cite{voi84}

Figure~\ref{fig:Growth_InPlane}(a) shows for excitation energies
between $1.535$~eV and $1.542$~eV an increase in $P_0$ as a function
of energy with a maximum of the measured $P_0 \approx 0.43$ at the
LH1:E1 transition ($1.542$~eV). Surprisingly, the measured $P_0$ at
the LH1:E1 transition is significantly lower than the calculated
value $P_0 = 0.6$. Such a difference between experiment and theory
is only observed for resonant in-plane excitation of the LH
transition. We discuss this point in more detail in
Sec.~\ref{sec:Bleaching}. For excitation energies above the LH
transition, $P_0$ is nearly constant apart from a dip at $1.608$~eV
which results from the HH2:E2 transition. Contrary to the case of
resonant HH1:E1 excitation, we do not obtain $P_0 \approx 0$ since
all excitons above the HH1:E1 absorption edge are Fano resonances
\cite{win95a} and the contribution of the LH1:E1 continuum gives
rise to a nonzero $P_0$.

\subsection{$P_0$ at the LH1:E1 resonance}
\label{sec:Bleaching}

\begin{figure}[t]
\includegraphics{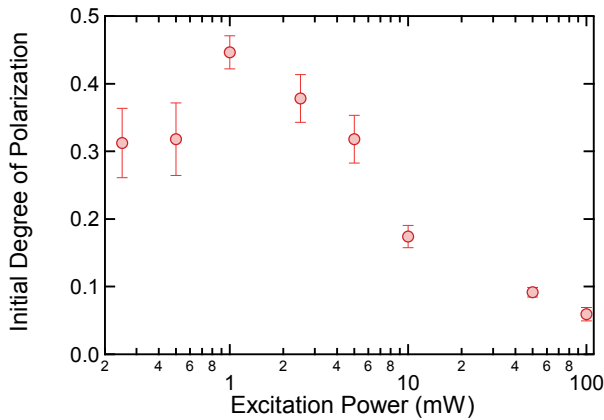}

\caption{\label{fig:Bleaching}(Color online) $P_0$ as a function of
 the excitation energy for resonant LH1:E1 excitation at $1.542$~eV.
 The excitation power of $1$~mW for maximum $P_0$ is comparable to
 the excitation power used in Fig.~\ref{fig:Growth_InPlane}.}
\end{figure}

In this section we discuss possible reasons for the reduction of the
measured $P_0$ at the LH1:E1 transition. At first glance,
spin-dependent phase-space filling of excitonic states could be a
possible explanation for this behavior. For sufficiently large
excitation powers $P_\mathrm{exc}$, the optically created electrons
and holes inhibit the creation of additional carriers due to Pauli
blocking. This effect is known as optical bleaching and leads to a
decrease in $P_0$ with increasing excitation powers. To check for
optical bleaching, we measure $P_0$ at the LH1:E1 transition as a
function of $P_\mathrm{exc}$ (see Fig.~\ref{fig:Bleaching}). The
measurements are performed with a nearly identical experimental
setup but, for experimental reasons, with a picosecond laser with a spectral
linewidth of $\approx 0.4$~nm FWHM. The experimental results clearly
show a strong influence of phase-space filling as $P_0$ decreases
with increasing $P_\mathrm{exc}$ for $P_\mathrm{exc} > 1$~mW.
However, we still measure a maximal $P_0 \approx 0.45$ since $P_0$
also decreases for $P_\mathrm{exc} < 1$~mW. Such a decrease of $P_0$
with decreasing excitation power has been observed before for
excitation in growth direction (see Fig.~6 in
Ref.~\onlinecite{pfa05}) and has been explained by a fast initial
spin relaxation, i.e., a fast excitonically induced electron spin
relaxation during the thermalization process which takes place
within our time resolution. We may expect that fast initial spin
relaxation is particularly important for the LH1:E1 resonance
because thermalization of the resonantly excited LH excitons
is much slower than the nearly instantaneous thermalization of
non-resonantly excited excitons in the continuum where
electron-electron and electron-phonon scattering occurs on very
short time scales of the order of 100~fs---consistent with the fact
that only for the LH1:E1 resonance do we measure a value of $P_0$ that
is significantly lower than theoretically expected.

The initial spin polarization $P_0$ may also be reduced due to an
enhanced broadening of the resonance lines for in-plane excitation
compared to excitation in growth direction. This can clearly be seen
by comparing the PLE data for both excitation geometries: The
resonances for in-plane excitation are shifted to lower energies as
in the case of excitaton in growth direction (see
Fig.~\ref{fig:Growth_InPlane}). It appears that this has a large
effect on the LH1:E1 transition, so that we might face an
energy-dependent line broadening here. Since the broadening of the
resonances (in our case approximately 3~meV) implies smaller values
for the measuerd $P_0$, our measured data for $P_0$ agrees well with
the measured PLE data for in-plane excitation.

\section{Theory}
\label{sec:theory}

To obtain a theoretical model for the optical spin orientation, it
is our goal to evaluate the spin density matrix for the optically
induced electron distribution. \cite{dym76} For clarity, we will
first develop the theory neglecting the Coulomb interaction between
electron and hole states. Then we will introduce the modifications
due to the formation of excitons.

\subsection{Single-particle spectrum}
\label{sec:theo-single}

The starting point of our theoretical discussion is an extension of
the general theory in Ch.~5 of Ref.\ \onlinecite{hau94} to
multicomponent single-particle states. \cite{win03} We describe the
system by means of the single-particle density operator $\rho$. We
use a basis of single-particle states (electrons and holes) of the
unperturbed system
\begin{equation}
  \label{eq:single-wf}
  \braket{\rr | \alpha\kk} \equiv
  \psi_{\alpha \kk} (\rr) = \sum_{n=1}^N
  \frac{e^{i \kk\cdot \vek{\rho}}}{2\pi} \; \xi_{\alpha \kk}^n (z)
  \: u_{n\vek{0}} (\rr) ,
\end{equation}
which are eigenfunctions of the $N\times N$ multiband Hamiltonian
$H_0$. Here $\rr = (\vek{\rho},z)$, $\kk$ is the wave vector for the
in-plane motion, and $u_{n\vek{0}} (\rr)$ are the basis functions of
$H_0$ which are Bloch functions for $\kk = \vek{0}$. The
position-dependent expansion coefficents are the spinors $e^{i
 \kk\cdot \vek{\rho}} \, \xi_{\alpha \kk}^n (z) / (2 \pi)$. The
energy eigenvalues corresponding to $\ket{\alpha\kk}$ are the
subband dispersions $E_\alpha (\kk)$, i.e., $H_0(\kk) \,
\ket{\alpha\kk} = E_\alpha (\kk)\, \ket{\alpha\kk}$. Now we can
write $\rho$ as
\begin{equation}
  \rho (\kk,t) = \sum_{\alpha,\alpha'}
  \rho_{\alpha,\alpha'} (\kk,t) \: \ket{\alpha\kk} \bra{\alpha'\kk}
\end{equation}
with expansion coefficients $\rho_{\alpha,\alpha'} (\kk,t)$. Here
the sums run over both the electron subbands $\alpha_e$ and the hole
subbands $\alpha_h$.
Using the dipole approximation, the light field is described by
\cite{scalar}
\begin{equation}
  \label{eq:light-ham}
  V = \frac{e}{m_0} A_0 \: \lim_{\eta \rightarrow 0}
  \left( e^{-i\omega t + \eta t} \: \hat{\vek{e}} \cdot \pp + \hrmcnjg \right)
  \equiv \frac{e}{m_0} \vek{A}_0 (t) \cdot \pp .
\end{equation}
Here $\vek{A}_0 (t)$ is the vector potential for the light field,
$\eta \rightarrow 0$ describes the adiabatic switching on of the
perturbation $V$, and $\hrmcnjg$ denotes the Hermitian conjugate of
the preceding term. We remark that for circularly polarized light,
the polarization vector $\hat{\vek{e}}$ is complex. The first term
in Eq.\ (\ref{eq:light-ham}) proportional to $e^{-i\omega t}$
describes absorption, while the second term describes emission. To
simplify our formulas, we neglect below all terms related with
emission.
Assuming that the envelope functions $e^{i \kk\cdot \vek{\rho}} \;
\xi_{\alpha \kk}^n (z)/(2\pi)$ are slowly varying on the length
scale of the lattice constant, we obtain for the matrix elements of
$\pp$ evaluated between electron and hole states
\begin{subequations}
  \label{eq:dipol-mat}
  \begin{eqnarray}
    \vekc{P}_{\alpha_h \alpha_e} (\kk)
    & = & \braket{\alpha_h \kk | \pp | \alpha_e\kk} \\
    & = & \sum_{n_e, \, n_h} \int d z \;
    \xi^{n_h \,\ast}_{\alpha_h \kk} (z) \:
    \xi^{n_e}_{\alpha_e \kk} (z) \:
    \braket {u_{n_h} | \pp | u_{n_e}} . \nonumber\\[-2ex]
  \end{eqnarray}
\end{subequations}
We neglect matrix elements of $\pp$ in between electron states and
in between hole states which would give rise to intraband optical
transitions in the infrared. Then $V$ becomes
\begin{equation}
  V (\kk,t) = \frac{e}{m_0}
  \sum_{\alpha_h,\alpha_e} 
  \Bigl[ \vek{A}_0 (t) \cdot \vekc{P}_{\alpha_h \alpha_e} (\kk) \;
   \ket{\alpha_h \kk} \bra{\alpha_e \kk} + \hrmcnjg \Bigr] .
\end{equation}

In the presence of the perturbation $V$, the density operator
$\rho(\kk,t)$ obeys the Liouville equation
\begin{equation}
  \label{eq:liou}
  \frac{d}{dt} \rho(\kk,t) =
  \frac{i}{\hbar} \left[\rho(\kk,t), H_0(\kk) + V(\kk) \right] .
\end{equation}
Switching to the interaction picture (superscript $I$), this
equation becomes
\begin{widetext}
\begin{equation}
  \begin{array}[b]{>{\displaystyle}r>{\displaystyle}l}
    \frac{d}{dt} \rho^I(\kk,t) = &
    \frac{ie}{\hbar \,m_0} \sum_{\alpha,\alpha'} \sum_{\alpha_h,\alpha_e}
    \rho_{\alpha,\alpha'}^I (\kk,t) \: \vek{A}_0 (t)
    \\ & \cdot
    \Bigl[ \; \vekc{P}_{\alpha_h \alpha_e} (\kk) \;
    e^{i[E_{\alpha_h} (\kk) - E_{\alpha_e} (\kk)]t/\hbar} \left(
    \ket{\alpha\kk} \braket{\alpha'\kk | \alpha_h \kk} \bra{\alpha_e \kk}
    - \ket{\alpha_h\kk} \braket{\alpha_e\kk | \alpha \kk} \bra{\alpha' \kk}
    \right)
    \\ & \hspace{0.3em}
    + \vekc{P}_{\alpha_e \alpha_h} (\kk) \;
    e^{i[E_{\alpha_e} (\kk) - E_{\alpha_h} (\kk)]t/\hbar} \left(
    \ket{\alpha\kk} \braket{\alpha'\kk | \alpha_e \kk} \bra{\alpha_h \kk}
    - \ket{\alpha_e\kk} \braket{\alpha_h\kk | \alpha \kk} \bra{\alpha' \kk}
    \right)\Bigr] .
  \end{array}
\end{equation}
As we neglected in Eq.\ (\ref{eq:dipol-mat}) the momentum matrix
elements in between electron states and in between hole states, this
equation can be decomposed into separate equations for the electron,
hole, and the off-diagonal electron-hole subspaces
\begin{subequations}
  \label{eq:dens-eom}
  \begin{eqnarray}
  \frac{d}{dt} \rho_{\alpha_e,\alpha_e'}^I (\kk,t) & = &
  \frac{ie}{\hbar \,m_0} \vek{A}_0 (t) \cdot \sum_{\alpha_h}
  \Bigl[ \hspace{0em} \vekc{P}_{\alpha_h \alpha_e'} (\kk) \:
  e^{-i[E_{\alpha_e'} (\kk) - E_{\alpha_h} (\kk)]t/\hbar}
  \rho_{\alpha_e,\alpha_h}^I (\kk,t)
  - \vekc{P}_{\alpha_e \alpha_h} (\kk) \:
  e^{i[E_{\alpha_e} (\kk) - E_{\alpha_h} (\kk)]t/\hbar}
  \rho_{\alpha_h,\alpha_e'}^I (\kk,t) \Bigr]
  \nonumber\\[-2ex] \label{eq:dens-eom-e} \\
  \frac{d}{dt} \rho_{\alpha_h,\alpha_h'}^I (\kk,t) & = &
  \frac{ie}{\hbar \,m_0} \vek{A}_0 (t) \cdot \sum_{\alpha_e}
  \Bigl[ \hspace{0em} \vekc{P}_{\alpha_e \alpha_h'} (\kk) \:
  e^{i[E_{\alpha_e} (\kk) - E_{\alpha_h'} (\kk)]t/\hbar}
  \rho_{\alpha_h,\alpha_e}^I (\kk,t)
  - \vekc{P}_{\alpha_h \alpha_e} (\kk) \:
  e^{-i[E_{\alpha_e} (\kk) - E_{\alpha_h} (\kk)]t/\hbar}
  \rho_{\alpha_e,\alpha_h'}^I (\kk,t) \Bigr]
  \nonumber\\[-2ex] \\
  \frac{d}{dt} \rho_{\alpha_e,\alpha_h}^I (\kk,t) & = &
  - \frac{ie}{\hbar \,m_0} \vek{A}_0 (t) \cdot
  \vekc{P}_{\alpha_e \alpha_h} (\kk) \:
  e^{i[E_{\alpha_e} (\kk) - E_{\alpha_h} (\kk)]t/\hbar}
  \left[ \rho_{\alpha_h,\alpha_h}^I (\kk,t)
    - \rho_{\alpha_e,\alpha_e}^I (\kk,t) \right] .
  \label{eq:dens-eom-eh}
  \end{eqnarray}
\end{subequations}

In quasi equilibrium (steady state) we can solve these equations
iteratively. \cite{hau94} To lowest order, the diagonal elements
$\rho_{\alpha,\alpha} (\kk,t)$ are given by thermal (Fermi)
distribution functions
\begin{equation}
  \rho_{\alpha,\alpha}^{(0)} (\kk)= f_\alpha (\kk) .
\end{equation}
To simplify the analysis we assume temperature $T=0$, i.e.,
$f_{\alpha_h} (\kk) = 1$ and $f_{\alpha_e} (\kk) = 0$. Now we can
integrate Eq.\ (\ref{eq:dens-eom-eh}) to obtain the block
off-diagonal elements of $\rho^I$
\begin{equation}
  \label{eq:dens-offd}
    \rho_{\alpha_e,\alpha_h}^I (\kk,t) =
    - \frac{e A_0}{\,m_0} \lim_{\eta \rightarrow 0}
      \frac{e^{i[E_{\alpha_e} (\kk) - E_{\alpha_h} (\kk)
        - \hbar\omega - i\hbar\eta ]t/\hbar}}
      {E_{\alpha_e} (\kk) - E_{\alpha_h} (\kk) - \hbar\omega - i\hbar\eta}
     \; \hat{\vek{e}} \cdot \vekc{P}_{\alpha_e \alpha_h} (\kk) .
\end{equation}
We insert Eq.\ (\ref{eq:dens-offd}) into Eq.\ (\ref{eq:dens-eom-e}).
Integrating the resulting equation and going back to the
Schr\"odinger picture we get \cite{dym76}
\begin{equation}
  \rho_{\alpha_e,\alpha_e'} (\kk,t) =
  \frac{e^2 A_0^2}{m_0^2} \sum_{\alpha_h} \;
  \lim_{\eta \rightarrow 0} \:
  \frac{[\hat{\vek{e}} \cdot \vekc{P}_{\alpha_e \alpha_h} (\kk)] \,
   [\hat{\vek{e}}^\ast \cdot \vekc{P}_{\alpha_h \alpha_e'} (\kk)]
   \; e^{2\eta t}}
  {[E_{\alpha_e} (\kk) - E_{\alpha_h} (\kk) - \hbar\omega + i\hbar\eta] \,
   [E_{\alpha_e'} (\kk) - E_{\alpha_h} (\kk) - \hbar\omega - i\hbar\eta]} .
\end{equation}
For the resonant case $E_{\alpha_e} (\kk) - E_{\alpha_h} (\kk) =
\hbar\omega$ the limit $\eta \rightarrow 0$ is ill-defined. Yet we
have
\begin{equation}
  \frac{d}{dt} \; \rho_{\alpha_e,\alpha_e} (\kk,t)
  \equiv \dot{\rho}_{\alpha_e,\alpha_e} (\kk,t)
  = \frac{2\pi}{\hbar} \, \frac{e^2 A_0^2}{m_0^2} \sum_{\alpha_h}
  \left|\hat{\vek{e}} \cdot \vekc{P}_{\alpha_e \alpha_h} (\kk)\right|^2 \;
  \delta [E_{\alpha_e} (\kk) - E_{\alpha_h} (\kk) - \hbar\omega]
\end{equation}
and the (dimensionless) absorption coefficient reads
\begin{subequations}
  \label{eq:absorb}
  \begin{eqnarray}
    \alpha (\omega) & = &
    \frac{e^2}{4\pi\epsilon_0\,\hbar c} \,
    \frac{4\pi}{n} \, \frac{m_0}{e^2A_0^2\, \omega}
    \trace \dot{\rho} (\omega) \\[2ex]
    & = &  \frac{\alpha_0}{\hbar\omega \, m_0}
    \sum_{\alpha_e, \alpha_h}
    \int\! \frac{d^2 k}{(2\pi)^2} \:
  \left|\hat{\vek{e}} \cdot \vekc{P}_{\alpha_e \alpha_h} (\kk)\right|^2 \;
  \delta [E_{\alpha_e} (\kk) - E_{\alpha_h} (\kk) - \hbar\omega] ,
  \end{eqnarray}
\end{subequations}
where the trace runs over all electron states and
\begin{equation}
  \alpha_0 \equiv \frac{e^2}{4\pi\epsilon_0\,\hbar c} \, \frac{8\pi^2}{n}
\end{equation}
with $\epsilon_0$ the permittivity of free space, $c$ the speed of
light and $n$ the index of refraction.

Finally, we obtain for the matrix elements $\varrho_{n_e, n_e'}
(\omega)$ of the electron spinor density matrix $\varrho$
\begin{equation}
  \label{eq:spinmat-single}
  \varrho_{n_e, n_e'} (\omega) =
  \frac{e^2 A_0^2}{m_0^2}
  \sum_{\alpha_e, \alpha_e'} 
  \int\! d z_e \;
  \xi^{n_e \,\ast}_{\alpha_e \kk} (z_e) \: \xi^{n_e'}_{\alpha_e' \kk} (z_e)
  \sum_{\alpha_h}
  \int\! \frac{d^2 k}{(2\pi)^2} \:
  \frac{[\hat{\vek{e}} \cdot \vekc{P}_{\alpha_e \alpha_h} (\kk)] \,
   [\hat{\vek{e}}^\ast \cdot \vekc{P}_{\alpha_h \alpha_e'} (\kk)]
   \; e^{2\eta t}}
  {[E_{\alpha_e} (\kk) - E_{\alpha_h} (\kk) - \hbar\omega + i\hbar\eta] \,
   [E_{\alpha_e'} (\kk) - E_{\alpha_h} (\kk) - \hbar\omega - i\hbar\eta]}
\end{equation}
and the spin polarization induced by a steady-state optical
excitation is given by
\begin{equation}
  \label{eq:spinpol}
  \vek{P} (\omega)
  = \frac{\trace \left[\vek{S} \, \varrho (\omega)\right]}
  {\trace \varrho (\omega)}
  = \frac{\trace \left[\vek{S} \, \dot{\varrho} (\omega)\right]}
  {\trace \dot{\varrho} (\omega)} \:,
\end{equation}
i.e., while the density matrices $\rho$ and $\varrho$ are
ill-defined in the limit $\eta \rightarrow 0$, the expectation
values of observables are well-defined and independent of $t$. (Yet
we keep $\eta >0 $ to simulate a finite broadening of the energy
levels.) The second equality in Eq.\ (\ref{eq:spinpol}) indicates
that $\vek{P} (\omega)$ is, indeed, equivalent to the definition of
the spin polarization proposed previously. \cite{pfa05} The
generalized spin matrices $\vek{S}$ are defined in Eq.\ (6.65) of
Ref.~\onlinecite{win03}.

\subsection{Excitonic spectrum}
\label{sec:theo_exc}

We can extend the theory developed in the previous subsection to
take into account the Coulomb interaction between electron and hole
states, thus giving rise to the formation of excitons. For this
purpose, we use the accurate exciton theory described in Ref.\
\onlinecite{win95a}. It is based on an axial approximation for
$H_0$, so that total angular momentum $l$ is a good quantum number.
We expand the exciton states in terms of the single-particle states
(\ref{eq:single-wf})
\begin{equation}
\Psi_{l\gamma} (\vek{r}_e, \vek{r}_h)
 =  \frac{1}{(2\pi)^{3/2}}
\sum_{\alpha_e, \, \alpha_h} \sum_{n_e, \, n_h}
\int d^2 k \:
\phi_{l\gamma\, k}^{\alpha_e \alpha_h}
\: e^{i \kk \cdot (\vek{\rho}_e - \vek{\rho}_h)}
\: e^{i (l - M_{n_e} + M_{n_h}) \varphi} \:
\xi^{n_e}_{\alpha_e k} (z_e) \, u_{n_e\vek{0}} (\rr_e)
\left[ \xi^{n_h}_{\alpha_h k} (z_h) \, u_{n_h\vek{0}} (\rr_h) \right]^\ast
\end{equation}
with expansion coefficients $e^{i l \varphi} \phi_{l\gamma\,
 k}^{\alpha_e \alpha_h}$ and $\kk = (k,\varphi)$. Similar to Eq.\
(\ref{eq:spinmat-single}), we obtain for the matrix elements
$\varrho_{n_e, n_e'} (\omega)$ of the electron spinor density matrix
$\varrho$
\begin{subequations}
  \label{eq:spinmat-ex}
  \begin{equation}
    \varrho_{n_e, n_e'} (\omega) =
    \frac{e^2 A_0^2}{m_0^2} \:
    \sum_{l, l'}
    \sum_{\gamma, \gamma'} \:
    \frac{[\hat{\vek{e}} \cdot \vekc{P}_{l\gamma}^\ast] \,
          [\hat{\vek{e}}^\ast \cdot \vekc{P}_{l'\gamma'}] \; e^{2\eta t}}
    {[E_{l\gamma} - \hbar\omega + i\hbar\eta] \,
     [E_{l'\gamma'} - \hbar\omega - i\hbar\eta]}
    \sum_{\alpha_e, \alpha_e'}
    \sum_{\alpha_h} \int\! d k \, k \;
    \phi_{l\gamma\, k}^{\alpha_e \alpha_h \ast} \:
    \phi_{l'\gamma'\, k}^{\alpha_e' \alpha_h} \:
    \int\! d z_e \;
    \xi^{n_e \,\ast}_{\alpha_e k} (z_e) \: \xi^{n_e'}_{\alpha_e' k} (z_e) \:,
  \end{equation}
  where
  \begin{equation}
    \vekc{P}_{l\gamma} =
    \sqrt{\frac{\mathcal{A}}{2\pi}}
    \sum_{\alpha_e, \, \alpha_h} \sum_{n_e, \, n_h}
    \int\! d k \, k \;
    \phi_{l\gamma\, k}^{\alpha_e \alpha_h} \:
    \delta_{l - M_{n_e} + M_{n_h}, \, 0} \;
    \int\! d z \;
    \xi^{n_h \,\ast}_{\alpha_h k} (z) \: \xi^{n_e}_{\alpha_e k} (z) \:
    \braket{u_{n_h} | \pp \, | u_{n_e}}
  \end{equation}
\end{subequations}
\end{widetext}
are the dipole matrix elements of the exciton states with
$\mathcal{A}$ the area of the QW interface. Once again, the
absorption coefficient is given by Eq.\ (\ref{eq:absorb}) (note
$\trace \rho = \trace\varrho$), and the optically induced spin
polarization is given by Eq.\ (\ref{eq:spinpol}). These equations
describe optical absorption and the resulting spin polarization for
arbitrary polarization directions $\hat{\vek{e}}$ of the exciting
light field.

\subsection{Kane Model}

For all numerical calculations presented in this work we have used
for the multiband Hamiltonian $H_0$ the $8\times 8$ Kane Hamiltonian
for the lowest conduction band $\Gamma_6^c$, the topmost valence
band $\Gamma_8^v$ and the spin split-off valence band $\Gamma_7^v$
including remote-band contributions of second order in $k$. This
Hamiltonian has been discussed in detail, e.g., in Ref.\
\onlinecite{win03}. It is known to provide an accurate description
of all important details of the semiconductor band structure
including the nonparabolic dispersion and the mixing of HH and LH
states. \cite{win95a, pfa05} The numerical values for all band
structure parameters were likewise taken from Ref.\
\onlinecite{win03}. The calculations were carried out using the
nominal growth parameters without any fitting parameters. The tiny
differences in energetic positions of the peaks in the measured and
calculated spectra in Fig.\ \ref{fig:Growth_InPlane} probably result
from the uncertainty in the Al concentration of the QW barriers.

Within the Kane model, we readily obtain Eq.\
(\ref{eq:spinmat-simple}) from Eq.\ (\ref{eq:spinmat-single}) if we
neglect the in-plane dispersion of the electron and hole states
(i.e., in-plane wave vector $\kk = 0$) as well as the $\kdotp$
coupling between conduction and valence band states so that all
subband states are represented by their dominant spinor component.

\subsection{Discussion}

The calculated spectra presented in
Figs.~\ref{fig:Growth_InPlane}(b) and (d) based on the excitonic
model of Sec.~\ref{sec:theo_exc} are overall in good agreement with
the measured spectra in Figs.~\ref{fig:Growth_InPlane}(a) and (c).
The simplified model from Sec.~\ref{sec:qual} neglecting Coulomb
coupling and HH-LH mixing suggests that individual peaks in the
spectra can be attributed to pairs of one electron and one hole
subband. As discussed in more detail in Refs.~\onlinecite{pfa05} and
\onlinecite{win95a}, such an approximate model is best justified for
the discrete excitonic states below the excitonic continuum. In the
excitonic continuum the excitons become Fano resonances \cite{fan61}
that are strongly modified by Coulomb coupling and HH-LH mixing of
individual subbands. These effects are immediately visible in the
calculated spectrum of Figs.~\ref{fig:Growth_InPlane}(b) (right
column). Here the spin polarization associated with the discrete
HH1:E1 exciton is close to zero, as expected based on the model of
Sec.~\ref{sec:qual} (being nonzero only because of the finite
broadening of the LH1:E1 exciton). On the other hand, the dip of the
polarization at the energy of the HH2:E2 exciton is essentially
independent of the phenomenological broadening, but it reflects the
finite width of a Fano resonance. \cite{win95a, fan61}

\section{Conclusion}
\label{sec:Conclusions}

The comparison of the results for excitation in the growth direction and
in-plane excitation clearly shows the dependence of the optical
selection rules on the excitation and detection geometry. Instead of
$P_0 = 1$ for excitation in the growth direction at the HH1:E1
resonance, we find $P_0 \approx 0$ for excitation in $y$-direction.
Moreover, we find no traces of a sign reversal in our data, i. e.,
there is no region with $P_0 \approx 0$ except for resonant HH1:E1
excitation. Such a sign reversal is typical for excitation in
the growth direction with energies close to the LH1:E1 resonance.

The experiments are well described by an accurate model for the spin
density matrix induced by the optical excitation. This model takes
into account both the effects of the semiconductor band structure
such as HH-LH coupling and nonparabolicity, and the Coulomb
coupling between electron and hole states giving rise to the
formation of excitons.

\begin{acknowledgments}
  The authors thank J.~Reno from Sandia National Lab for the
  excellent sample. The work was supported by the BMBF, the German
  Science Foundation (DFG--Priority Program 1285 ``Semiconductor
  Spintronics''), and the Centre for Quantum Engineering and
  Space-Time Research in Hannover (QUEST).
\end{acknowledgments}


\begin{thebibliography}{10}
\bibitem{scalar}
Throughout we define the scalar product as $\vek{a} \cdot \vek{b} = \sum_i a_i
  b_i$.
\end{thebibliography}
\end{document}